\documentclass[12pt,a4paper]{article}
\usepackage{graphicx}

\makeatletter
\newcounter{lineno}
\def\verbatimlisting#1{\setcounter{lineno}{0}%
    \begingroup{\footnotesize} \@verbatim \frenchspacing \@vobeyspaces
\parindent=20pt
    \everypar{\stepcounter{lineno}\llap{\thelineno\ \ }}\input#1
    \endgroup
}
\makeatother

\newcommand{\cu}{{\cal U}}
\newcommand{\cv}{{\cal V}}
\newcommand{\um}{{\bar u}}
\newcommand{\gm}{\mbox{gm}}
\newcommand{\secs}{\mbox{s}}
\newcommand{\cm}{\mbox{cm}}
\newcommand{\Ord}[1]{{\cal O}\left(#1\right)}
\newcommand{\rr}{{\sf R}}

\begin{document}

\title{\sf Advection-dispersion in symmetric field-flow fractionation
channels}
\author{S.A. Suslov and A.J. Roberts
\thanks{Dept of Maths \& Comput, University of
Southern Queensland, Toowoomba, Queensland 4350, \textsc{Australia}.
E-mail: \texttt{ssuslov@usq.edu.au, aroberts@usq.edu.au} respectively.}}
\maketitle

\begin{abstract}
We model the evolution of the concentration field of
macro\-mole\-cules in a symmetric field-flow fractionation (FFF)
channel by a one-di\-men\-sion\-al advection-diffusion equation.  The
coefficients are precisely determined from the fluid dynamics.  This
model gives quantitative predictions of the time of elution of the
molecules and the width in time of the concentration pulse.  The model
is rigorously supported by centre manifold theory.  Errors of the
derived model are quantified for improved predictions if necessary.
The advection-diffusion equation is used to find that the optimal
condition in a symmetric FFF for the separation of two species of
molecules with similar diffusivities involves a high rate of
cross-flow.
\end{abstract}

\tableofcontents

\section{Introduction}
\label{Sintro}

\begin{figure}[tbp]
  \centerline{\includegraphics[width=0.9\textwidth]{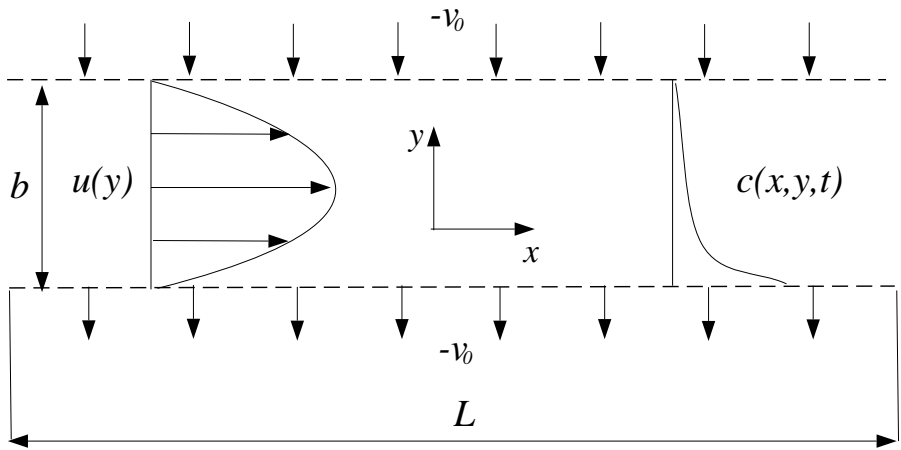}}
  \caption{Side view of symmetric field-flow fractionation (FFF) channel.}
  \label{f1}
\end{figure}

Consider the transport of some contaminant molecules in the fluid flow
of a symmetric field-flow fractionation (FFF) channel as analysed by
Giddings and others \cite[e.g.]{Giddings86,Schure89} and sketched in
Figure\ref{f1}.
The two horizontal parallel plates above and below the channel are not
permeable to the contaminant molecules but allow for the cross-flow of
fluid.
This cross-flow distributes the contaminant preferentially to the
lower side of the channel as shown in Figure~\ref{f2}.
It is this cross-flow and asymmetric distribution of contaminant
concentration $c(x,y,t)$ that creates a differential advection of
different molecular species and renders the problem interesting.

Using techniques based upon centre manifold theory \cite{Roberts88a},
from the continuum equations (Section~\ref{Seqns}) we deduce that a
model for the contaminant distribution in the channel is the
advection-diffusion equation
\begin{equation}
    \frac{\partial C}{\partial t}=-U\frac{\partial C}{\partial x}
    +D\frac{\partial^2 C}{\partial x^2}\,,
     \label{Emod}
\end{equation}
where $t$ denotes time, $x$ measures distance downstream along the
channel, and $C(x,t)=c(x,0,t)$ is the concentration of the contaminant
measured along the lower plate (the so called accumulating wall).
We derive expressions for the effective advection velocity $U$ as it
predominantly determines the time of efflux of the contaminant out
across the end of the channel, and the effective diffusivity $D$ as it
determines how wide the contaminant spreads by the time it reaches the
end of the channel: in a useful parameter regime
(Section~\ref{Sdetail})
\begin{equation}
	U\approx \frac{6\bar u\kappa}{v_0b}\,,\quad
	D\approx\frac{72\bar u^2\kappa^3}{v_0^4b^2}\,,
	\label{Eudap}
\end{equation}
where $\kappa$ is the molecular diffusivity, $\bar u$ is the mean
along-channel velocity, $b$ is the channel height, and $v_0$ is the
cross-flow velocity.
The term $D\frac{\partial^2 C}{\partial x^2}$ models the so called
``zone broadening effects'' discussed by Litzen and others
\cite{Litzen90, Schure89}.
We also quantify the two sources of errors in the model by
\begin{itemize}
  \item estimating the time it takes for initial transients to die out
  and the model to become valid (Section~\ref{Sdappr});

  \item determining higher-order corrections to the advection-diffusion
  model (Section~\ref{Sdetail}).
\end{itemize}
This model and its errors may be rigorously justified as discussed in
other applications of centre manifold theory to shear dispersion by
Mercer, Roberts and Watt
\cite{Roberts88a,Mercer90,Mercer94a,Watt94c,Watt94b}.

Field-flow fractionation channels are used to separate species of
contaminant molecules with different diffusivities.
In Section~\ref{Ssep} we use model~(\ref{Emod}) to identify that FFF
separates molecular species most efficiently for relatively high
cross-flow: up to
\begin{equation}
	v_0\approx 6^{3/4}\sqrt{\frac{\bar u\kappa}{b}}\,.
	\label{Evap}
\end{equation}
Consequently, in describing the governing equations in
Section~\ref{Seqns} we introduce a non-dimensionalisation appropriate
for such high cross-flow rates.

Further research in field-flow fractionation could model the
dynamics of contaminant molecules in tubular channels \cite{Wahlund87},
trapezoidal channels \cite{Litzen90}, or in asymmetric FFF channels
\cite{Litzen93}, as well as the dynamics of non-neutrally buoyant
particles \cite{Schure89}.

\section{Governing equations for symmetric FFF}
\label{Seqns}

Consider a symmetric FFF channel as discussed by Giddings and others
\cite{Giddings86,Schure89,Wyatt98} and depicted schematically in
Figure~\ref{f1}.
The dynamics takes place between two flat plates located at $y=0$ and
$y=b$.
The fluid flow between the plates is driven predominantly by a
pressure gradient $p_x$ parallel to the plates.
Being that of a Newtonian fluid with kinematic viscosity $\nu$ and
density $\rho$, the velocity field is essentially that of parabolic
Poiseuille flow except that there is a cross-flow, of velocity $-v_0$,
from the upper plate to the lower (if $v_0$ is positive).
The plates are permeable to the fluid in order for this cross-flow to
occur; but they are impermeable to the contaminant molecules.
Within the fluid the contaminant, of concentration $c(x,y,t)$, is
advected by the flow and diffuses with coefficient $\kappa$.
In this section we non-dimensionalise the governing differential
equations, and also deduce the advecting fluid velocity field and
confirm that it is nearly parabolic.

For order of magnitude estimates of quantities we use the geometry of
Wahlund \& Giddings \cite{Wahlund87}: the channel width is
$b\approx0.05\,\cm$; the density of the fluid, water, is
$\rho=1\,\gm/\cm^3$; and the kinematic viscosity $\nu\approx
0.01\,\cm^2/\secs$.
The fluid moves so that on average it takes about 5--15~minutes to
traverse about $50\,\cm$ so a typical fluid velocity is $\um\approx
0.1\,\cm/\secs$ and the driving pressure gradient must be roughly
$p_x\approx 5\,\gm/\cm^2/\secs^2$.
The cross-flow is driven at rates $v_0$ of order
$5\times10^{-4}\,\cm/\secs$.
When the contaminant molecules are the Cow Pea Mosaic Virus
\cite[p464]{Litzen93}, this configuration gives parameters as listed
in Table~\ref{Tparam}.
We base our analysis on this set being typical of parameters of
interest.
\begin{table}[tbp]
  \centering
	\caption{Typical set of physical parameters for FFF and the
	consequent parameters (in the second part) appearing in the
	analysis.  The data is for the Cow Pea Mosaic Virus
	\cite[p464]{Litzen93} in the FFF channel of \cite{Wahlund87}.}
  \begin{tabular}{|lc|c|}
    \hline
    Parameter & & Value  \\
    \hline\hline
    Channel width & $b$ & $0.05\,\cm$  \\
    \hline
    Kinematic viscosity & $\nu$ & $0.01\,\cm^2/\secs$  \\
    \hline
    Mean longitudinal velocity & $\um$ &$0.1\,\cm/\secs$  \\
    \hline
    Cross-flow velocity & $v_0$ & 5$\times10^{-4}\,\cm/\secs$  \\
    \hline
    Molecular diffusivity & $\kappa$  &$2\times10^{-7}\,\cm^2/\secs$  \\
    \hline\hline
    Boundary layer (BL) thickness &$\eta$& $4\times10^{-4}\,\cm$\\
    \hline
    Cross-BL diffusion time & $\tau$ & $0.8\,\secs$  \\
    \hline
    Longitudinal velocity in the BL&$u_0$&$5\times10^{-3}\,
    \cm/\secs$\\
    \hline
    Downstream advection distance & $\xi$ & $4\times 10^{-3}\,\cm$\\
    \hline
    Prandtl number & $\sigma$ &$5\times10^4$  \\
    \hline
    Cross-channel Peclet number & $\cv$ &  $125$  \\
    \hline
    Downstream Peclet number& $\cu$&$2.5\times10^4$\\
    \hline
    Velocity ratio & ${\cal K}$ &  $0.1$  \\
    \hline
  \end{tabular}
  \label{Tparam}
\end{table}

The equations governing the fluid motion are the Navier-Stokes and
continuity equations
\begin{equation}
  \frac{\partial{\bf q}}{\partial t}+{\bf q}\cdot\nabla{\bf q}=
  -\frac{1}{\rho}\nabla p+\nu\nabla^2{\bf q}\,,
  \label{Enavs}
\end{equation}
\begin{equation}
  \nabla\cdot{\bf q}=0
  \label{Ecty}
\end{equation}
for the incompressible velocity field ${\bf q}=u{\bf i}+v{\bf j}$ and
for the pressure $p$.  The contaminant evolves according to the
advection--diffusion equation
\begin{equation}
        \frac{\partial c}{\partial t}+{\bf q}\cdot\nabla c=\kappa\nabla^2c
\end{equation}
for the concentration field $c$.
Herein we assume the molecules are neutrally buoyant, though
sedimentation effects \cite{Schure89} could be included in further
work by modifying this equation.
Note that although we are concerned with the dynamics of the
concentration field $c$, we only seek the steady and $x$-independent
fluid flow governed by the Navier-Stokes and continuity equations.
The boundary conditions on the plates are those of no longitudinal
flow,
\begin{equation}
        u=0\,,\quad v=-v_0\,,\quad\mbox{on $y=0$ and $y=b$}\,,
\end{equation}
and no flux of the contaminant through the plates,
\begin{equation}
		v_0c+\kappa\frac{\partial c}{\partial y}=0\,,\quad\mbox{on
		$y=0$ and $y=b$}\,.
\end{equation}
The above equations fully specify the dynamics of the fluid and the
contaminant molecules in the channel.

The non-dimensionalisation we adopt is chosen to reflect the fact that
for the regime of most effective separation of species (see
Section~\ref{Ssep}) the contaminant is concentrated near the lower
plate due to the cross-flow.  Introduce the following non-dimensional
variables denoted by stars:
\begin{equation}
        x^*=\frac{x}{\xi}\,,\quad
        y^*=\frac{y}{\eta}\,,\quad
        t^*=\frac{t}{\tau}\,,\quad
        u^*=\frac{u}{u_0}\,,\quad
        v^*=\frac{v}{v_0}\,,\quad
        p^*=\frac{p}{\rho v_0^2}\,,
\end{equation}
where $\eta=\kappa/v_0$ is the characteristic thickness of the
distribution of contaminant in a boundary layer near the lower plate,
$\tau=\eta/v_0=\eta^2/\kappa=\kappa/v_0^2$ is the cross-boundary layer
advection ($\eta/v_0$) or equivalently the cross-boundary layer
diffusion ($\eta^2/\kappa$) time,
\begin{equation}
   u_0=-\frac{1}{2}\frac{\partial p}{\partial x}\frac{b\eta}{\rho\nu}=
\frac{6{\um}}{\cv},
\end{equation}
is the characteristic downstream velocity in the boundary layer,
${\um}=-\frac{1}{12}\frac{\partial p}{\partial x}\frac{b^2}{\rho\nu}$
is the mean speed of the Poiseuille flow in absence of the cross-flow,
$\xi=u_0\tau$ is the downstream advection distance for the
material in the boundary layer in a cross-boundary layer diffusion
time, and where
\begin{equation}
  \sigma=\frac{\nu}{\kappa} \quad \mbox{and} \quad \cv=\frac{v_0b}{\kappa}
\end{equation}
are Prandtl and cross-channel Peclet numbers, respectively.
Typical values of all these quantities are recorded in
Table~\ref{Tparam}.
In essence this scaling is that of the distribution of contaminant
molecules which typically are swept to be near the lower plate with
the upper plate ``far away'' at $y=\cv$.
Substitute these scalings into the equations and omit the
distinguishing stars hereafter.

The steady fluid flow is straightforward to determine.  The
$y$-momentum equation determines that $v=-1$ everywhere.  The
$x$-momentum equation for the steady velocity field $u(y)$ becomes
\begin{equation}
        \frac{\cv}{2}\left[\frac{1}{\sigma}\frac{\partial u}{\partial y}
        +\frac{\partial^2u}{\partial y^2}\right]+1=0
\end{equation}
with boundary conditions $u(0)=u(\cv)=0$. The exact
solution for this velocity component is
\begin{eqnarray}
  u(y)&=&\frac{2\sigma}{\cv}\left[\cv\frac{1-e^{-y/\sigma}}
    {1-e^{-\cv/\sigma}}-y\right]\\
  &=&\left(y-\frac{y^2}{\cv}\right)
    +\frac{y}{\sigma}\left(\frac{\cv}{6}-\frac{y}{2}+\frac{y^2}{3\cv}
    \right)+\Ord{\frac{\cv^2}{\sigma^2}}\,.
\end{eqnarray}
Observe, as used in earlier analyses \cite[e.g.]{Wahlund87}, the
downstream advection is nearly parabolic; because the Prandtl number
$\sigma$ is so large the correction of $\Ord{\cv/\sigma}$ is usually
negligible.

The dynamics of the contaminant remains nontrivial.  Under our
nondimensionalisation the advection-diffusion equation becomes
\begin{equation}
    \frac{\partial c}{\partial t}+u\frac{\partial c}{\partial x}
    -\frac{\partial c}{\partial y}=\frac{\partial^2 c}{\partial y^2}
    +{\cal K}^2\frac{\partial^2 c}{\partial x^2}\,,
\label{ade}
\end{equation}
where
\begin{equation}
  {\cal K}=\frac{v_0}{u_0}=\frac{\cv}{6}\frac{v_0}{\um}=\frac{\cv^2}{6\cu}
  \quad \mbox{and} \quad \cu=\frac{\um b}{\kappa}
\end{equation}
are respectively the velocity ratio and the downstream Peclet number
based on the mean longitudinal speed, see Table~\ref{Tparam}.  The
non-dimensional boundary conditions for the contaminant are
\begin{equation}
  c+\frac{\partial c}{\partial y}=0\quad \mbox{at $y=0$ and $y=\cv$}\,.
\label{Eadebc}
\end{equation}
We analyse the dynamics described by this non-dimensional equation in
this paper.  The main non-dimensional parameter $\cv$, appearing as
the non-dimensional width of the channel, is typically large,
$\Ord{10^2}$, as we expect cross-flow advection to keep the
contaminant close to the bottom plate.

\section{The dynamics approach a centre manifold}
\label{Sdappr}

We justify the basis of model~(\ref{Emod}) using centre manifold
theory \cite{Carr81} as adapted \cite{Roberts88a,Mercer90} to the long
thin geometry of the FFF channel.  Under the action of the cross-flow
balanced by diffusion the contaminant distribution across the channel
relaxes quickly to an exponential distribution, $c=C\exp(-y)$.  The
shear velocity, different at different $y$, will smear this
contaminant cloud out along the channel, while cross-flow and
diffusion continue to act to push the cross-channel distribution
towards the exponential distribution.  The net effect is that the
cloud has a concentration that is slowly varying along the channel and
is approximately exponential across it.  Thus, after the quick decay
of cross-stream transients, we justify the relatively slow long-term
evolution of a contaminant cloud for which $x$ derivatives of $C$,
$\partial^nC/\partial x^n$, are small.

An initial ``linear'' picture of the dynamics is established by
assuming that there are no downstream variations.  When downstream
gradients are ignored, the relaxation across the channel of the
contaminant obeys the dynamics
\begin{equation}
        \frac{\partial c}{\partial t}-\frac{\partial c}{\partial y}=
        \frac{\partial^2 c}{\partial y^2}\,,
        \quad\mbox{s.t.}\quad
        c+\frac{\partial c}{\partial y}=0\mbox{ on $y=0$ and $y=\cv$}\,.
\end{equation}
The neutral solution already mentioned is the exponential
$c_0=C\exp(-y)$.  The other solutions, all decaying, are
\begin{equation}
  c_n=C_n\left[\sin\left(\frac{n\pi}{\cv}y\right) -\frac{2n\pi}{\cv}\cos
    \left(\frac{n\pi}{\cv}y\right)
  \right]\exp\left( -\frac{y}{2}+\lambda_n t \right)\,,
\end{equation}
for $n=1,2,\ldots$, where $C_n$ are constant coefficients determined
by the initial condition such that
\begin{equation}
  c(y,0)=\sum_{n=0}^\infty c_n(y,0) \label{icc}
\end{equation}
and the decay rate is
\begin{equation}
        \lambda_n=-\frac{1}{4}-\frac{n^2\pi^2}{\cv^2}\,.
        \label{Elamb}
\end{equation}

The slowest rate of decay to the centre manifold will be due to the
$n=1$ mode, although in many cases the second term in~(\ref{Elamb}) is
negligibly small as $\cv$ is of order $10^2$.
As an example, assume an initially uniform distribution of contaminant
across the channel, then from~(\ref{icc}) expect that $C_1\propto
\exp(\cv/2)$.
The relaxation process is then dominated by the exponential decay of
$\exp(\cv/2+\lambda_1t)$ which effectively leads to a relaxation time
of roughly $t_{\mbox{\scriptsize rel}}=-\cv/(2\lambda_1)$.
In dimensional form this cross-channel relaxation time
\begin{displaymath}
        t_{\mbox{\scriptsize rel}}\approx 2\cv\tau=200\,\secs\,,
\end{displaymath}
which agrees with the experimental observations of several minutes for
macromolecules given, for example, in \cite[Eqn~(30)]{Wahlund87}.
Thus expect the decay to a low-dimensional centre manifold to occur on
this time scale.

The presence of downstream $x$-variations perturbs the contaminant
pulse and results in its non-trivial long-time evolution.  Centre
manifold theory provides a powerful rationale for modelling such
evolution where the long-term behaviour is separated from rapidly
decaying transients.  This was recognised by Coullet \& Spiegel
\cite{Coullet83} and Carr \& Muncaster \cite{Carr83a,Carr83b}; see the
draft review by Roberts \cite{Roberts97a} for an extensive discussion.
The application of the theory to dispersion in channels and pipes has
been developed by Roberts, Mercer and Watt
\cite{Roberts88a,Mercer90,Mercer94a,Watt94b,Watt94c}.  Using the same
techniques here, we seek a solution to the governing equations in the
form
\begin{equation}
        c=h(C,y)\quad\mbox{such that}\quad
        \frac{\partial C}{\partial t}=g(C)\,.
        \label{Ecm}
\end{equation}
Here the function $h$, $C\exp(-y)$ to leading approximation, describes
the details of the contaminant field throughout space and time in
terms of the concentration $C$ of contaminant at the lower plate.
Such a solution forms a model of the dynamics for two reasons.  First,
the low-dimensional set of states described by $h(C)$ are
exponentially attractive because of the action of cross-stream
advection and diffusion as seen above.  Secondly, the associated
function $g$ models the effective advection and diffusion of the
contaminant in the horizontal by describing the evolution of $C$.

We find approximations to these functions by assuming that the
concentration field is slowly varying in the horizontal, that is,
$\partial/\partial x$ is a small operator.
Rigorously, one would expand in the downstream wavenumber as
introduced by Roberts~\cite{Roberts88a}.
Formally we express $h$ and $g$ in the following asymptotic series
\begin{equation}
        g\sim\sum_{n=1}^\infty g_n\frac{\partial^nC}{\partial x^n}
        \quad\mbox{and}\quad
        h\sim\sum_{n=0}^\infty h_n(y)\frac{\partial^nC}{\partial x^n}\,,
        \label{Easym}
\end{equation}
where for example $h_0=\exp(-y)$ is the leading order approximation to
the contaminant field, $-g_1=U$ is the effective advection velocity,
and $g_2=D$ is an effective horizontal diffusion coefficient.  The
advection-diffusion model~(\ref{Emod}) is obtained from just the first
two terms in the expansion for $g$.  In dispersion problems, the
asymptotic series in~(\ref{Easym}) typically converge in a sense
discussed by Mercer, Roberts and Watt \cite{Mercer90,Mercer94a,Watt94b}.

To find the asymptotic expansions~(\ref{Easym}) we implement an
iterative algorithm (see \cite{Roberts96a}) in computer algebra (see
Appendix~\ref{Scap}).
The results are assured to be accurate by the approximation theorem of
centre manifolds.
Assume that some approximate solution of the contaminant
advection-diffusion equation~(\ref{ade}) with boundary
conditions~(\ref{Eadebc}) is found in the centre manifold
form~(\ref{Ecm}); for example, the iteration is initiated with the
approximation $c=C\exp(-y)$ and $g=0$.
We wish to refine such an approximation by finding a correction
$h^\prime$ to the shape of the centre manifold and a correction $g'$
to the evolution thereon.
As established by Roberts \cite{Roberts96a} the corrections are found
by solving
\begin{equation}
  \frac{\partial^2 h'}{\partial y^2}+\frac{\partial h'}{\partial y}
  =\rr+g'\exp(-y)\,,
  \label{Ecmeq}
\end{equation}
where $\rr$ is the residual of~(\ref{ade}), with boundary conditions
\begin{equation}
   h'+\frac{\partial h'}{\partial y}=0\quad\mbox{at $y=0$ and $y=\cv$,}
   \quad\mbox{and}\quad
   h'=0\quad\mbox{at $y=0$.}
   \label{Ecmbc}
\end{equation}
This last boundary condition reflects that we seek a solution
parameterized by the concentration at the lower plate:
$C(x,t)=c|_{y=0}$.  The correction to the evolution $g'$ is chosen to
satisfy the solvability condition
\begin{equation}
  \int_0^\cv \rr+g'\exp(-y)\,dy=0
\end{equation}
in order to satisfy boundary conditions~(\ref{Ecmbc}).  Then the
differential equation~(\ref{Ecmeq}) is solved to find $h'$.  The
iterations continue until the desired terms are found in the
asymptotic approximation to the centre manifold~(\ref{Easym}).
Computer algebra, such as the program listed in Appendix~\ref{Scap},
easily performs all the algebraic details.

\section{The detailed centre manifold model}
\label{Sdetail}

Since all the algebraic machinations are handled by the computer
algebra of Appendix~\ref{Scap}, here we just record and discuss the
results.  General results simplify considerably in the typical
case of large $\cv$ when the contaminant is held near the lower
plate.  Then higher order corrections are readily found.

From the computer algebra results, the concentration field (the centre
manifold) is to low order
\begin{eqnarray}
  c&=&Ce^{-y}\nonumber\\
  & &{}-\frac{\partial C}{\partial x}\left(1-\frac{1}{\sigma}\right)
  \left[2me^{-\cv}-e^{-y}\left(2me^{-\cv}(y+1)-\frac{y^2}{2}+\frac{y^2}{\cv}
      +\frac{y^3}{3\cv}\right)\right]\nonumber\\
  & &{}-\frac{1}{12\sigma}\frac{\partial C}{\partial x}\frac{e^{-y}y^2}{\cv}
  (\cv-y)^2+\Ord{\frac{\partial^2 C}{\partial x^2},
    \frac{\cv^2}{\sigma^2}}\,,
\label{ccmf}
\end{eqnarray}
where the evolution of the contaminant concentration along the bottom
plate is described to leading order by
\begin{equation}
  \frac{\partial C}{\partial t}=\frac{\partial C}{\partial x}
  \left[\left(1-\frac{1}{\sigma}\right)\left(1-2m+\frac{2}{\cv}\right)
    -\frac{\cv}{6\sigma}\right]
  +\Ord{\frac{\partial^2 C}{\partial x^2},\frac{\cv^2}{\sigma^2}}\,,
\end{equation}
where
\begin{equation}
	m=\left(1-e^{-\cv}\right)^{-1}
	\sim\left\{
	\begin{array}{ll}
		1 & \mbox{as $\cv\to\infty$}  \\
		1/\cv+1/2 & \mbox{as $\cv\to0$}
	\end{array}
	 \right.\,.
	\label{Emdef}
\end{equation}
The order of error notation $\Ord{\alpha,\beta}$ is used to denote
errors $\Ord{\alpha}+\Ord{\beta}$.

\begin{figure}[tbp]
  \centerline{\includegraphics[width=0.9\textwidth]{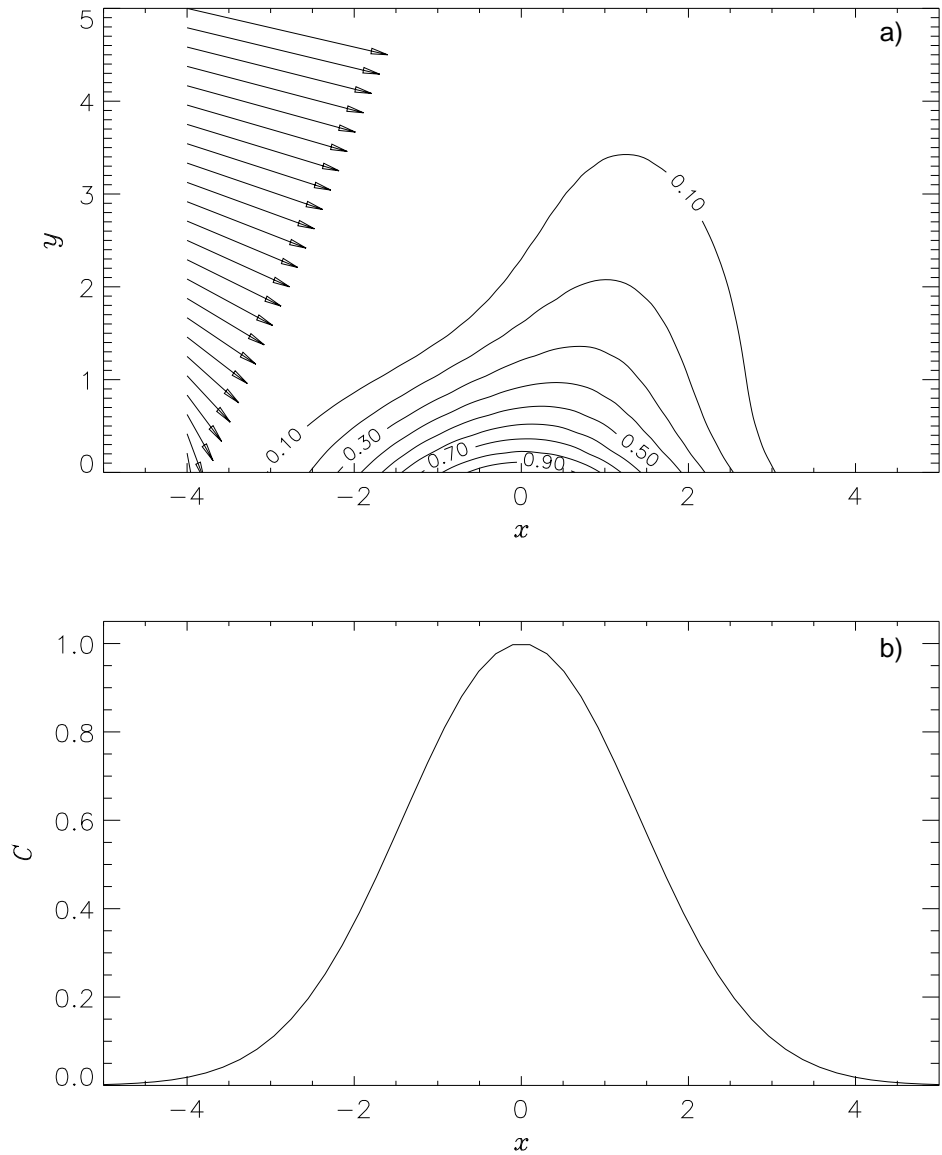}}
  \caption{Velocity field and an instantaneous concentration field near
  the accumulating lower plate (a) when the concentration along the wall is
  given by the Gaussian (b).  Fields correspond to the parameters given
  in Table~\ref{Tparam}.}
  \label{f2}
\end{figure}

Since the typical cross-channel Peclet number $\cv$ is of order $10^2$
we take $m=1$ in presenting further detailed results (for completeness
we present results for weak and moderate cross-flows in
Appendix~\ref{Sweak}).
The dominant error in this approximation is $\Ord{e^{-\cv}}$ and so
expect it to be acceptable for $\cv$ greater than about~$6$.
Then (\ref{ccmf}) simplifies to
\begin{eqnarray}
  c&=&Ce^{-y}+\frac{\partial C}{\partial x}y^2
    \left[\left(1-\frac{1}{\sigma}\right)
    \left(\frac{3+y}{3\cv}-\frac{1}{2}\right)
    -\frac{(\cv-y)^2}{12\cv\sigma}\right]e^{-y}\nonumber\\
  & &{}+\Ord{\frac{\partial^2 C}{\partial x^2},\frac{\cv^2}{\sigma^2},
    e^{-\cv}}\,.
\label{conc}
\end{eqnarray}
This shows the predominantly exponential distribution of the
contaminant as advection towards the lower plate by the cross-flow is
counter balanced by diffusion.
The exponential distribution is modified by the interaction of the
shear flow and the along-channel spatial gradients of the contaminant
as given by the second term in~(\ref{conc}) and shown in
Figure~\ref{f2}.
The above expressions give the details of the concentration field
parameterized by its value $C(x,t)=c|_{y=0}$ at the lower plate.

The associated advection-diffusion equation is~(\ref{Emod})
with coefficients
\begin{eqnarray}
  U&=& \left(1-\frac{2}{\cv}\right)+\frac{1}{\sigma}
  \left(\frac{\cv}{6}-1+\frac{2}{\cv}\right)+\Ord{\frac{\cv^2}{\sigma^2},
    e^{-\cv}}\,,\label{Eu}\\
  D&=&\left({\cal K}^2+2-\frac{20}{\cv}+\frac{56}{\cv^2}\right)
  +\frac{2(\cv-8)}{3\sigma}\left(1-\frac{12}{\cv}+\frac{42}{\cv^2}\right)
  \nonumber\\&&{}
  +\Ord{\frac{\cv^2}{\sigma^2}, e^{-\cv}}
\label{Ed}
\end{eqnarray}
giving the effective advection speed and dispersion coefficient.  The
crudest approximation, but useful over a reasonable parameter regime,
is that $U\approx1$ and $D\approx2$ leading to the dimensional
expressions given in the Introduction.

Running the computer algebra program to higher order in spatial
gradients we find that the dynamics of the dispersion is governed by
the extended evolution equation
\begin{equation}
  \frac{\partial C}{\partial t}=-U\frac{\partial C}{\partial x}
  +D\frac{\partial^2 C}{\partial x^2}
  +E\frac{\partial^3 C}{\partial x^3}
  +F\frac{\partial^4 C}{\partial x^4}
  +\Ord{\frac{\partial^5 C}{\partial x^5}}\,,
  \label{Emode}
\end{equation}
where the coefficients of the third and fourth order derivatives are:
\begin{eqnarray}
  E & = & -4\left(5-\frac{102}{\cv}+\frac{744}{\cv^2}
  -\frac{1936}{\cv^3}\right)
  \label{Ese}  \\
  & &{}-\frac{2}{\sigma}\left(5\cv-170+\frac{2220}{\cv}-\frac{13408}{\cv^2}+
    \frac{31856}{\cv^33}\right)+\Ord{\frac{\cv^2}{\sigma^2},e^{-\cv}}\,,
  \nonumber\\
  F & = & 16\left(22-\frac{725}{\cv}+\frac{9480}{\cv^2}-\frac{58292}{\cv^3}+
    +\frac{142168}{\cv^4}\right)  \nonumber\\
  & &{}+\frac{16}{3\sigma}\left(44\cv-2175
    +\frac{44145}{\cv}-\frac{464560}{\cv^2}+\frac{2549556}{\cv^3}
    -\frac{5856960}{\cv^4}\right)\nonumber\\
  & &{}+\Ord{\frac{\cv^2}{\sigma^2},e^{-\cv}}\,.\label{Esf}
\end{eqnarray}
The $\partial^3_x$ term in (\ref{Emode}) with coefficient $E$ governs
the skewness of the predictions of the model by modifying the
effective advection speed of various spatial modes.
The $\partial^4_x$ term with coefficient $F$ affects the decay of the
spatial modes.
Note that $F$ is positive for at least large enough $\cv$ and
$\sigma$---a fourth order model may thus be unstable for short enough
spatial modes: approximately the fourth order model is unstable for
along channel non-dimensional wavenumbers $|k|>1/(4\sqrt{11})$.
Thus although the third-order term may be used to improve predictions
of the advection-diffusion model, the fourth-order term should be
limited to helping estimate errors in the predictions.

\section{Species separate best at high cross-flow}
\label{Ssep}

The aim of field-flow fractionation is to separate as far as possible
two or more different species of contaminant molecules.
Different contaminants are characterised by different diffusivities,
$\kappa_j$ say.
A contaminant with lower diffusivity will be pushed closer to the
lower plate by the cross-flow.
Consequently, its effective advection speed along the channel will be
lower.
Thus one collects a contaminant with higher diffusivity at the exit
before a contaminant with lower diffusivity.
Here we identify the operating regime when the separation is most
effective between two species of nearly the same diffusivity.

\begin{figure}[tbp]
  \centerline{\includegraphics[width=0.9\textwidth]{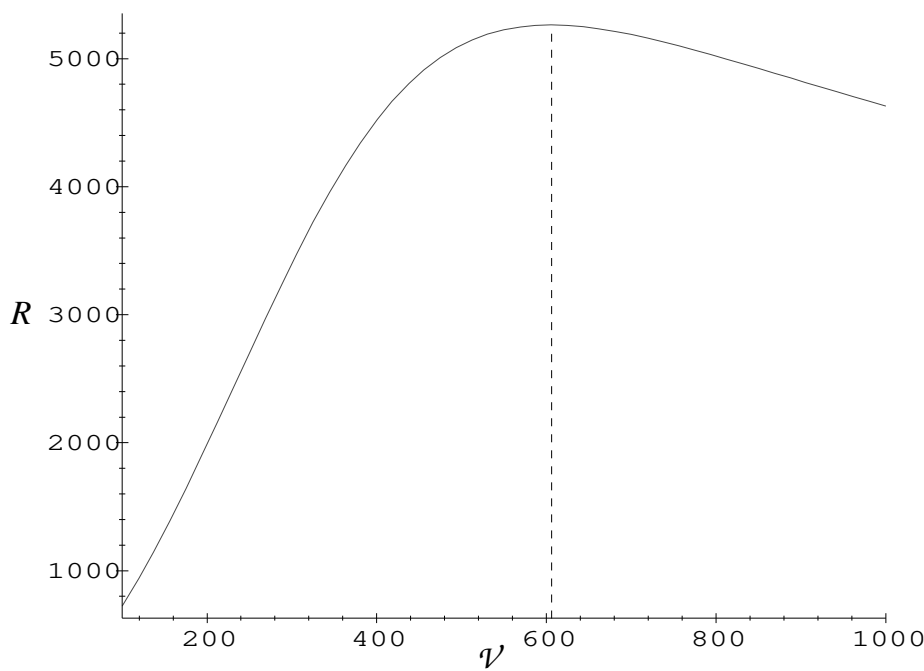}}
  \caption{Function $R(\cv)$ characterising effectiveness of
  separation of two different contaminants for parameters given in
  Table~\ref{Tparam}.}
  \label{R_P}
\end{figure}

Consider the advection-diffusion predicted by model~(\ref{Emod}) with
different species identified by the subscript $j$.
In the non-dimensional analysis this leads to different characteristic
scales:
\begin{equation}
  \tau_j=\frac{\kappa_j}{v_0^2}\,,\quad
  \xi_j=u_{0j}\tau_j
  =\frac{6\bar u\kappa_j^2}{v_0^3b}\,,\quad
  \cv_j=\frac{v_0b}{\kappa_j}\,,\quad
  \sigma_j=\frac{\nu}{\kappa_j}\,.
  \label{Escalj}
\end{equation}
Thus the advection-diffusion model~(\ref{Emod}) for the
$j$th species has dimensional coefficients
\begin{equation}
  U_j=u_{0j} U(\cv_j)\,,\quad D_j=\frac{\xi_j^2}{\tau_j}D(\cv_j)\,,
  \label{Edimc}
\end{equation}
from the leading term in each of~(\ref{Eu})--(\ref{Ed}) upon neglecting
terms of order $\cv_j/\sigma_j$.  In a channel of fixed length $L$ the
approximate times of efflux are
\begin{equation}
  T_j=\frac{L}{U_j}=\frac{L}{u_{0j}U(\cv_j)}
  =\frac{L}{6{\um}}\frac{\cv_j}{U(\cv_j)}\,.
  \label{Etout}
\end{equation}
Then the time interval between the moments when the two contaminant
pulses with diffusivities $\kappa_1=\kappa-\Delta\kappa/2$ and
$\kappa_2=\kappa +\Delta\kappa/2$, injected simultaneously at the
beginning of the channel, exit the channel is
\begin{equation}
  \Delta T\approx\left|\frac{\partial T}{\partial\kappa}\Delta\kappa\right|=
  \frac{P}{6}\frac{L}{{\um}}\frac{|U(\cv)-\cv U^\prime(\cv)|}{U^2(\cv)}
    \frac{\Delta\kappa}{\kappa}\,.
\end{equation}
The width of the contaminant pulse at the time of efflux is
proportional to $\sqrt{DT}$, and hence the time taken for a
contaminant pulse to pass the end is proportional to $\delta$ where
\begin{equation}
  \delta^2=\frac{DT}{U^2}
  =L\frac{\tau^2}{\xi}\frac{D(\cv)}{U^3(\cv)}
  =\frac{1}{6}\frac{L}{{\um}}\frac{b^2}{\kappa \cv}\frac{D(\cv)}{U^3(\cv)}\,.
  \label{Etwid}
\end{equation}
To maximise separation of two species with close values of diffusion
coefficients $\kappa_j$ we need to maximise the difference in the time
of efflux relative to the width in time of the pulses at the efflux.
Thus for a given small change in diffusivity, $\Delta\kappa>0$, we
wish to maximise
\begin{equation}
  \frac{\Delta T}{\delta}=
  \left|\frac{\partial T}{\partial \cv}
  \frac{\partial \cv}{\partial \kappa}\right| \frac{\Delta\kappa}{\delta}
  =\frac{\cv}{\delta}
  \left|\frac{\partial T}{\partial \cv}\right|
  \frac{\Delta\kappa}{\kappa}
  =\sqrt{\frac{L}{6b\cu }}
  \frac{\cv^{3/2}|U(\cv)-\cv U'(\cv)|}{\sqrt{D(\cv)U(\cv)}}
  \frac{\Delta\kappa}{\kappa}\,.
        \label{Esj}
\end{equation}
Expect the existence of an optimum cross-flow from the following
physical arguments.
Increasing the cross-flow significantly increases the difference in
efflux times.
On other hand, an extremely strong cross-flow would keep both
contaminants close to the plate in a very slow flow for long enough so
that longitudinal molecular diffusion becomes significant.
Thus resolution will decrease for an excessively strong cross-flow.
The optimal separation of species with given diffusivities in a
channel of fixed geometry with a fixed fluid flux through it is
accomplished when
\begin{eqnarray}
  R(\cv)&=&\frac{\cv^{3/2}|U(\cv)-\cv U'(\cv)|}{\sqrt{D(\cv)U(\cv)}}
  \nonumber\\
  &=&\frac{6\cu \cv^2(\cv-4)}{\sqrt{(\cv-2)
  \left(\cv^6+36 \cu^2(2\cv^2-20\cv+56)\right)}}
   \label{Erj}
\end{eqnarray}
is maximised. From $dR/d\cv=0$ we obtain
\begin{equation}
\cv^6\left(\cv^2-12\cv+16\right)=72\cu^2
\left(3\cv^4+52\cv^3-336\cv^2+912\cv-896\right)
\end{equation}
with a solution for optimal $\cv$ of
\begin{equation}
  \cv_0=6^{3/4}\sqrt{\cu}+\frac{22}{3}
  -\frac{353}{27}\frac{6^{1/4}}{\sqrt{\cu}}
  +\Ord{\frac{1}{\cu}}\,.
\end{equation}
The leading term of this optimum gives the optimum $v_0$ mentioned in
the Introduction.  As seen from Figure~\ref{R_P}, for the parameter
values listed in Table~\ref{Tparam} this optimum occurs at
$\cv_0\approx613$ which corresponds to the relatively high cross-flow
velocity $v_0\approx2.5\times10^{-3} \cm/\secs$.
Then the optimal regime of two species separation gives
\begin{eqnarray}
  \frac{\Delta T}{\delta}
  &=&\frac{6^{1/8}}{2}\sqrt{\frac{3L}{b}}\frac{\Delta\kappa}{\kappa}\cu^{1/4}
  \nonumber\\
  &&{}\times\left[1
  -6^{1/4}\frac{7}{24}\cu^{-1/2}
  -\frac{1997}{384}\sqrt{\frac{2}{3}}\cu^{-1}
  +\Ord{\cu^{-3/2}}\right]\,.
\end{eqnarray}
For the geometry of the channel considered in \cite{Wahlund87} and
parameters given in Table~\ref{Tparam} the maximum resolution is thus
\begin{equation}
  \frac{\Delta T}{\delta}\approx430\frac{\Delta\kappa}{\kappa}\,,
\end{equation}
where $\delta\approx1$\,min.  For the regime considered the time
necessary for the contaminant to travel a distance $L=50\,\cm$ is
$T\approx14.2$\,hours---probably too long to be practical.  A
suggestion is to reduce the channel length or increase the
longitudinal flow speed, while increasing the cross-flow velocity to
be closer to the optimum.

\paragraph{Acknowledgement} We thank the Australian Research Council
for a grant supporting this research, and Dr~Bob Anderssen of CSIRO
for introducing us to this problem.

\appendix

\section{Computer algebra handles all the details}
\label{Scap}

Just one of the virtues of this centre manifold approach to modelling
is that it is systematic.  This enables relatively straightforward
computer programs to be written to find the centre manifold and the
evolution thereon \cite[e.g.]{Roberts96a}.

For this problem the iterative algorithm is implemented by a computer
algebra program written in \textsc{reduce} \footnote{At the time of
writing, information about \texttt{reduce} was available from Anthony
C.~Hearn, RAND, Santa Monica, CA~90407-2138, USA.
E-mail: \tt reduce@rand.org} Although there are many details in the
program, the correctness of the results are \emph{only determined} by
driving to zero (line~48) the residual of the governing differential
equation, evaluated on line~41, to the error specified on line~38 and
with boundary and amplitude conditions checked on lines~50--52.
The other details only affect the rate of convergence to the ultimate
answer.

{\footnotesize \verbatimlisting{sfffp.red}}

\section{Weak and moderate cross-flows}
\label{Sweak}

For completeness we record here the model for the case of relatively
slow cross-flow, or equivalently of relatively high diffusivity.  This
provides results for all parameters $\cv$, not just the large values
described earlier.

The non-dimensionalisation used in the main body of this paper is
inappropriate in the case of small cross-flow rates.  At higher rates
the contaminant is restricted to the boundary layer, but in low
cross-flow it is spread over the channel height.  Thus in the case of
weak cross-flows we adopt the following scalings typical of those
used for shear dispersion \cite[e.g.]{Mercer90,Watt94b}:
\begin{equation}
        y^*=\frac{y}{b}\,,\quad
        t^*=\frac{\kappa t}{b^2}\,,\quad
        x^*=\frac{x\kappa}{{\um}b^2}\,,\quad
        u^*=\frac{u}{{\um}}\,,\quad
        v^*=\frac{bv}{\kappa}\,,\quad
        p^*=\frac{p}{\rho\um^2\sigma}\,.
\label{nd1}
\end{equation}
Quantities are scaled: $y$ with the channel width; $t$ with a
cross-channel diffusion time, $\tau=b^2/\kappa\approx 1.25\times 10^4$
sec; $x$ with the downstream advection distance in a cross-channel
diffusion time, $\xi={\um}\tau\approx1.25\times 10^3$ cm; $u$ with the
mean downstream velocity; and $v$ with a cross-stream diffusion speed,
$\kappa/b\approx4\times10^{-4}$ cm/s.  As before $\cv=v_0b/\kappa$ is
the main parameter and is used to denote a non-dimensional cross-flow
velocity, though it may well be thought of as an effective channel
width, or as the inverse of the molecular diffusivity.

Then after substituting~(\ref{nd1}) into the Navier-Stokes and
continuity equations~(\ref{Enavs})--(\ref{Ecty}) and dropping stars
the equation for the steady horizontal velocity component $u(y)$
becomes
\begin{equation}
  -\frac{\cv}{\sigma}\frac{\partial u}{\partial y}=
  12+\frac{\partial^2 u}{\partial y^2}\,,
  \quad\mbox{s.t.}\quad
  u=0 \quad\mbox{at $y=0$ and $y=1$}
\end{equation}
with the nearly parabolic solution
\begin{eqnarray}
  u(y)&=&\frac{12\sigma}{\cv}
  \left(\frac{1-e^{-\cv y/\sigma}}{1-e^{-\cv/\sigma}}-y\right)\,
  \nonumber\\
  &=&y(1-y)\left(6+(1-2y)\frac{\cv}{\sigma}
  +\Ord{\frac{\cv^2}{\sigma^2}}\right)\,.
\end{eqnarray}

The advection-diffusion equation for the contaminant becomes
\begin{equation}
  \frac{\partial c}{\partial t}+u\frac{\partial c}{\partial x}
  -\cv\frac{\partial c}{\partial y}=\frac{\partial^2 c}{\partial y^2}
  +\frac{1}{\cu^2}\frac{\partial^2 c}{\partial x^2}\,,
  \label{Ect}
\end{equation}
where $\cu=\um b/\kappa$ is a downstream Peclet number as before, and
with boundary conditions
\begin{equation}
  \cv c+\frac{\partial c}{\partial y}=0\quad\mbox{at $y=0$ and $y=1$.}
\label{adebc1}
\end{equation}
In the absence of any $x$-variations the steady solution is
\begin{equation}
  c_0=Ce^{-\cv y}\,,
\label{c0b}
\end{equation}
where as before $C=c(x,0,t)$ is the concentration of the contaminant
at the lower plate.  The other $x$-independent solutions, all
decaying, are
\begin{equation}
        c_n=C_n\left[\cv\sin\left(n\pi y\right) -2n\pi\cos
        \left(n\pi y\right)
        \right]\exp\left( -\frac{\cv y}{2}+\lambda_n t \right)\,,
\end{equation}
for $n=1,2,\ldots$ where the decay rate is
\begin{equation}
        \lambda_n=-\frac{\cv^2}{4}-n^2\pi^2\,.
\end{equation}
For small $\cv$ the decay is dominated by the second term above due to
cross-channel diffusion.
The slowest rate of decay to the centre manifold comes from the $n=1$
mode.
Using arguments similar to those given in Section~\ref{Sdappr} we
deduce that in the case of small cross-flow rates $C_1\propto 1$ and,
consequently, the dimensional decay time is expected to be
$\tau/|\lambda_1| \approx\tau/\pi^2 \approx20$\,min which is an order
of magnitude larger than that for the strong cross-flows considered
earlier.
This reaffirms the existence of an attractive centre manifold for
slowly varying solutions, albeit attractive on a larger time scale.

As before, an iterative procedure was implemented in computer (not
listed) to solve the contaminant transport
equations~(\ref{Ect})--(\ref{adebc1}) by finding the centre manifold
and the evolution thereon~(\ref{Ecm}).  The resulting expression for
the concentration field is
\begin{eqnarray}
  c&=& C e^{-\cv y}+12\frac{\partial C}{\partial x}\frac{m-1}{\cv^3}
  \left(1-\frac{1}{\sigma}\right)\left(1-e^{-\cv y}(1+\cv y)\right)
  \nonumber\\& &{}
  +\frac{\partial C}{\partial x}\frac{y^2}{\cv^2}e^{-\cv y}
\left[(6-3\cv+2\cv y)\left(1-\frac{1}{\sigma}\right)
-\frac{\cv^2}{2\sigma}(1-y)^2\right]
\nonumber\\&&{}
+\Ord{\frac{\partial^2 C}{\partial x^2},\sigma^{-2}}\,.
\label{conc1}
\end{eqnarray}
The general expressions for the coefficients of the evolution
equation~(\ref{Emode}) are quite involved and for brevity here we
neglect terms inversely proportional to the Prandtl number $\sigma$
since it is typically small ($\sigma\approx 5\times10^4$ for example):
\begin{eqnarray}
  U&=& \frac{6}{\cv}(2m-1)-\frac{12}{\cv^2}
  +\Ord{\sigma^{-1}}\,,\\
  D&=&\frac{1}{\cu^2}+24m\frac{e^{-\cv}}{\cv^2}
  -144m^2(2m-1)\frac{e^{-\cv}}{\cv^3}
  -72\frac{4m^2e^{-\cv}-1}{\cv^4}
  \nonumber\\&&{}
  -720\frac{2m-1}{\cv^5}+\frac{2016}{\cv^6}
  +\Ord{\sigma^{-1}}\,,\\
  E & = & -24m^2(2m-1)\frac{e^{-\cv}}{\cv^3}
  +\frac{432}{5}m^2\left(20m^2e^{-\cv}+3\right)\frac{e^{-\cv}}{\cv^4}
  \nonumber\\&&{}
  -6912m^4(2m-1)\frac{e^{-2\cv}}{\cv^5}
  -24192m^4\frac{e^{-2\cv}}{\cv^6}
  \nonumber\\&&{}
  -864(2m-1)\frac{58m^2e^{-2\cv}+5}{\cv^7}
  -5184\frac{48m^2e^{-\cv}-17}{\cv^8}
  \nonumber\\&&{}
  -642816\frac{2m-1}{\cv^9}+\frac{1672704}{\cv^{10}}
  +\Ord{\sigma^{-1}}\,,
  \label{Ese1}  \\
  F & = & 16m^2\left(6m^2e^{-\cv}+1\right)\frac{e^{-\cv}}{\cv^4}
  -\frac{1152}{5}m^2(2m-1)\left(15m^2e^{-\cv}+1\right)\frac{e^{-\cv}}{\cv^5}
  \nonumber\\&&{}
  +\frac{288}{35}m^2(2m-1)\left(168m^2(100m^2e^{-\cv}+21)e^{-\cv}+37\right)
  \frac{e^{-\cv}}{\cv^6}
  \nonumber\\&&{}
  -1152m^2(2m-1)\left(24m^2\left(15m^2e^{-\cv}-1\right)e^{-\cv}-1\right)
  \frac{e^{-\cv}}{\cv^7}
  \nonumber\\&&{}
  -\frac{6912}{5}m^2\left(60m^2\left(25m^2e^{-\cv}-2\right)e^{-\cv}-79\right)
  \frac{e^{-\cv}}{\cv^8}
  \nonumber\\&&{}
  -6912m^2(2m-1)\left(630m^2e^{-\cv}+17\right)
  \frac{e^{-\cv}}{\cv^9}
  \nonumber\\&&{}
  -13824\frac{m^2\left(1806m^2e^{-\cv}+197\right)e^{-\cv}-33}
  {\cv^{10}}
  \nonumber\\&&{}
  -518400(2m-1)\frac{106m^2e^{-\cv}+29}{\cv^{11}}
  -829440\frac{463m^2e^{-\cv}-237}{\cv^{12}}
  \nonumber\\&&{}
  -1208742912\frac{2m-1}{\cv^{13}}+\frac{2947995648}{\cv^{14}}
  +\Ord{\sigma^{-1}}\,.
\label{Esf1}
\end{eqnarray}
\begin{figure}[tbp]
        \centerline{\includegraphics[width=0.9\textwidth]{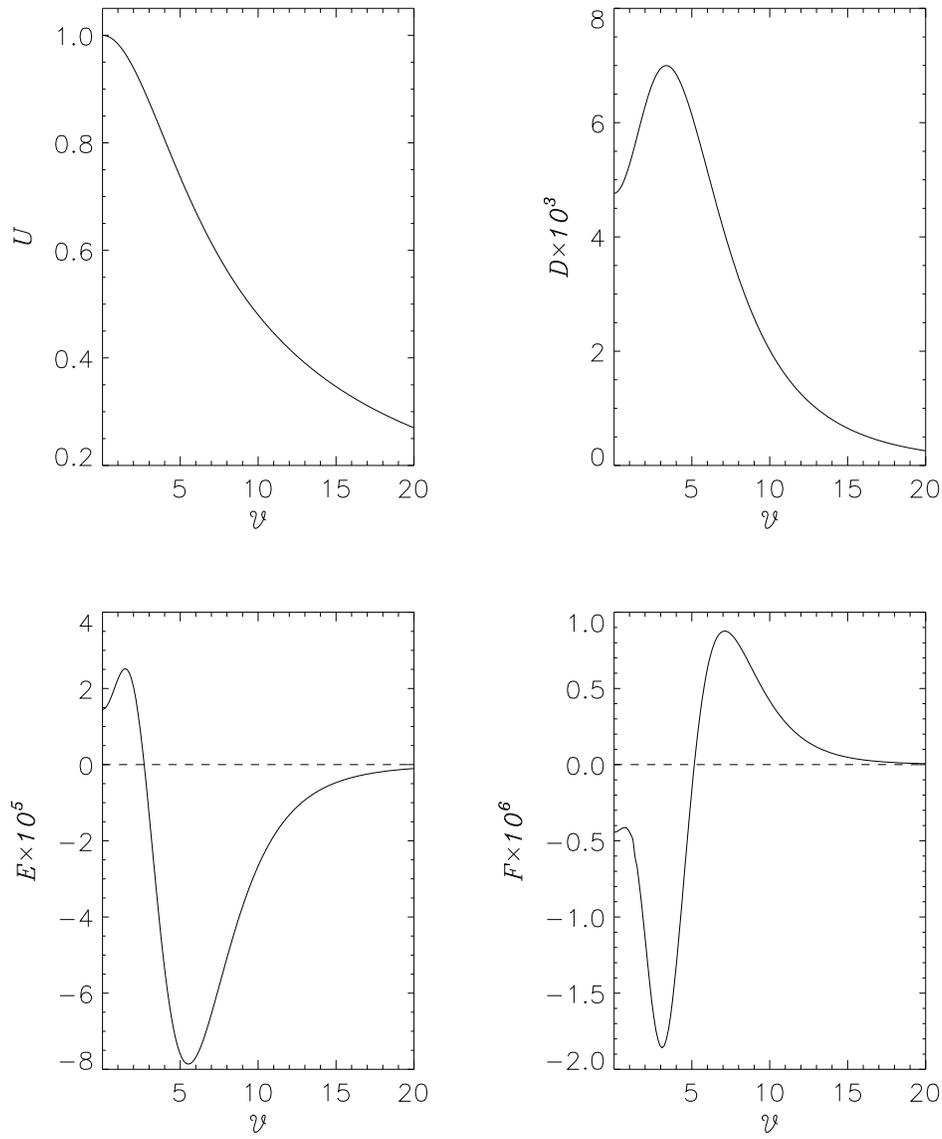}}
		  \caption{The coefficients of the evolution equation
		  (\ref{Emode}) as functions of the cross-channel Peclet
		  number $\cv$ for the infinite Prandtl number $\sigma$.}
        \protect\label{Feffc}
\end{figure}

These coefficients are plotted in Figure~\ref{Feffc}.  All of them
eventually decrease in magnitude with increasing cross-flow.  The
maximum of the effective diffusion coefficient $D$ is reached at
$\cv\approx3$.  Thus the ``zone broadening'' \cite{Litzen90}
associated with the dispersion of the contaminant affects the
distribution of the contaminant at a greater degree when the
cross-flow is relatively weak.  The fourth order coefficient $F$ is
negative for $\cv<5$ and consequently equation~(\ref{Emode}), in
contrast to the case of strong cross-flows reported in
Section~\ref{Sdetail}, predicts stable (decaying) in time evolution of
the average concentration of the contaminant for all longitudinal
wavenumbers.

The small $\cv$ expansions of expressions (\ref{conc1})--(\ref{Esf1})
are
\begin{eqnarray}
  c&=& C\left[1-\cv y+\frac{1}{2}\cv^2y^2-\frac{1}{6}\cv^3y^3\right]
+\frac{\partial C}{\partial x}y^2\left[-\frac{1}{2}(y-1)^2\right.
\nonumber\\&&
   {}+\frac{\cv y}{60}\left(24y^2-45y+20
     +\frac{1}{\sigma}\left(6y^2-15y+10\right)\right)
     \nonumber\\&&{}
     +\frac{\cv^2}{120}\left(20y^4-36y^3+15y^2-1
   -\frac{1}{\sigma}\left(10y^4-24y^3+15y^2+1\right)\right)
   \nonumber\\&&{}
   \left.+\frac{\cv^3 y}{1260}\left(60y^4-105y^3+42y^2-7
+\frac{1}{\sigma}\left(45y^4-105y^3+63y^2+7\right)\right)\right]
\nonumber\\&&
 {}+\Ord{\frac{\partial^2 C}{\partial x^2},\sigma^{-2},\cv^4}\,,
\\
  U&=&1-\frac{\cv^2}{60}+\frac{\cv^4}{2520}-\frac{\cv^6}{100800}
  +\Ord{\sigma^{-1},\cv^8}\,,\\
  D&=&\frac{1}{\cu^2}+\frac{1}{210}\left(1+\frac{7}{60}\cv^2
    -\frac{89}{7920}\cv^4+\frac{239}{386100}\cv^6\right)
  +\Ord{\sigma^{-1},\cv^8},\\
  E&=&\frac{1}{69300}\left(1+\frac{1073}{1365}\cv^2
    -\frac{233}{910}\cv^4+\frac{71629}{2570400}\cv^6\right)
  +\Ord{\sigma^{-1},\cv^8}\,,\\
  F&=&-\frac{1}{2252250}\left(1-\frac{26879}{85680}\cv^2
    +\frac{26341969}{68372640}\cv^4+\frac{1187149277}{15041980800}
	\cv^6\right)
  \nonumber\\&&{}+\Ord{\sigma^{-1},\cv^8}\,.
\end{eqnarray}
The expansions for large cross-flow $\cv$ are
\begin{eqnarray}
  c&=& Ce^{-\cv y}\nonumber\\
& &{}+\frac{\partial C}{\partial x}y^2e^{-\cv y}\left[
  -\frac{1}{2\sigma}(y-1)^2
  +\left(\frac{2y-3}{\cv}+\frac{6}{\cv^2}\right)
  \left(1-\frac{1}{\sigma}\right)\right]\nonumber\\
&&{}+\Ord{\frac{\partial^2 C}{\partial x^2},\sigma^{-2},e^{-\cv}}\,,
\\
  U&=&\frac{6}{\cv}\left(1-\frac{2}{\cv}\right)
  +\Ord{\sigma^{-1},e^{-\cv}}\,,\\
  D&=&\frac{1}{\cu^2}+\frac{72}{\cv^4}\left(1-\frac{10}{\cv}
    +\frac{28}{\cv^2}\right)
  +\Ord{\sigma^{-1},e^{-\cv}}\,,\\
  E&=&-\frac{4320}{\cv^7}\left(1-\frac{102}{5\cv}
  +\frac{744}{5\cv^2}-\frac{1936}{5\cv^3}\right)
  +\Ord{\sigma^{-1},e^{-\cv}}\,,\\
  F&=&\frac{456192}{\cv^{10}}\left(1
  -\frac{725}{22\cv}+\frac{4740}{11\cv^2}
  -\frac{29146}{11\cv^3}+\frac{71084}{11\cv^4}\right)
  +\Ord{\sigma^{-1},e^{-\cv}}.
\end{eqnarray}
The above expressions are equivalent to, but appear a little different
from, the leading terms in expressions (\ref{conc})--(\ref{Ed}),
(\ref{Ese}) and (\ref{Esf}) because of the different
non-di\-men\-sion\-al\-isation.
These large $\cv$ expressions evidently give the behaviour of the
coefficients for non-dimensional cross-flow $\cv$ bigger than about
5--10.

\addcontentsline{toc}{section}{\refname}
\bibliographystyle{plain}\bibliography{ajr,new,bib}

\end{document}